\documentclass[11pt]{article}
\usepackage{amsfonts}
\usepackage{a4,tikz}
\usepackage{geometry}
\usepackage{color} 
\usepackage{subfigure}
\makeatletter
\newcommand\ackname{Acknowledgements}
\if@titlepage
  \newenvironment{acknowledgements}{%
      \titlepage
      \null\vfil
      \@beginparpenalty\@lowpenalty
      \begin{center}%
        \bfseries \ackname
        \@endparpenalty\@M1
      \end{center}}%
     {\par\vfil\null\endtitlepage}
\else
  
\fi
\makeatother
\usepackage{amsmath}
\usepackage{amssymb}
\usepackage{amsthm}
\usepackage{mathrsfs}
\usepackage{eucal}
 \usepackage[usenames,dvipsnames]{pstricks}
 \usepackage{epsfig}
 \usepackage{pst-grad} 
 \usepackage{pst-plot} 
\renewcommand{\theequation}{\arabic{equation}}
\usepackage{dsfont}
\usepackage{amssymb,amsmath}
\usepackage{color}
\usepackage{accents}
\usepackage{texdraw}
\usepackage{eucal}
\usepackage{epic,epsfig}
\usepackage{graphicx}

\theoremstyle{definition}

\numberwithin{equation}{section}
\DeclareMathAccent{\wtilde}{\mathord}{largesymbols}{"65}
\DeclareMathAccent{\what}{\mathord}{largesymbols}{"62}

\def\m@th{\mathsurround=0pt}
\mathchardef\bracell="0365
\def\upbrall{$\m@th\bracell$}
\def\undertilde#1{\mathop{\vtop{\ialign{##\crcr
    $\hfil\displaystyle{#1}\hfil$\crcr
     \noalign
     {\kern1.5pt\nointerlineskip}
     \upbrall\crcr\noalign{\kern1pt
   }}}}\limits}
\def\m@th{\mathsurround=0pt}
\mathchardef\bracell="0365
\def\upbrall{$\m@th\bracell$}
\def\underhat#1{\mathop{\vtop{\ialign{##\crcr
    $\hfil\displaystyle{#1}\hfil$\crcr
     \noalign
     {\kern1.5pt\nointerlineskip}
     \upbrall\crcr\noalign{\kern1pt
   }}}}\limits}
\usepackage{pgf}
\usepackage{tikz}
\usepackage{verbatim}
\usepackage[utf8]{inputenc}
\usetikzlibrary{arrows,automata}
\usetikzlibrary{positioning}
\usepackage[toc,page]{appendix}

\def\theequation{\arabic{section}.\arabic{equation}}

\newcommand{\bblu}{\begin{color}{blue}}
\newcommand{\bred}{\begin{color}{red}}
\newcommand{\ecl}{\end{color}}

\newcommand{\be}{\begin{equation}}
\newcommand{\ee}{\end{equation}}
\newcommand{\bea}{\begin{eqnarray}}
\newcommand{\eea}{\end{eqnarray}}
\newcommand{\bse}{\begin{subequations}}
\newcommand{\ese}{\end{subequations}}

\title{The emergence of the relativistic Lagrangian from the non-relativistic multiplicative Lagrangian}
\author{Kittikun Surawuttinack$^{1,\dagger}$, Suppanat Supanyo$^{2,*}$ and  Sikarin Yoo-Kong$^{2,\ddagger}$ \\
\small {$^1$Department of Physics,}
\small \emph{ King Mongkut's University of Technology Thonburi, Bangkok, Thailand, 10400.}\\ 
            \small {$^2$The Institute for Fundamental Study (IF),} \small\emph{Naresuan University, Phitsanulok, Thailand, 65000.}\\
		\small{$^\dagger$Kittikun.Surawuttinack@gmail.com, $^*$suppanatsupanyo@gmail.com,$^\ddagger$sikariny@nu.ac.th} \\
	}
\begin{document}
\def\theequation{\arabic{section}.\arabic{equation}}

\newtheorem{thm}{Theorem}[section]
\newtheorem{lem}{Lemma}[section]
\newtheorem{defn}{Definition}[section]
\newtheorem{ex}{Example}[section]
\newtheorem{rem}{}
\newtheorem{criteria}{Criteria}[section]
\newcommand{\ra}{\rangle}
\newcommand{\la}{\langle}
\maketitle

\begin{abstract}
The multiplicative Lagrangian and Hamiltonian introduce an additional parameter that, despite its variation, results in identical equations of motion as those derived from the standard Lagrangian. This intriguing property becomes even more striking in the case of a free particle. By manipulating the parameter and integrating out, the statistical average of the multiplicative Lagrangian and Hamiltonian naturally arises. Astonishingly, from this statistical viewpoint, the relativistic Lagrangian and Hamiltonian emerge with remarkable elegance. On the action level, this formalism unveils a deeper connection: the spacetime of Einstein’s theory reveals itself from a statistical perspective through the action associated with the multiplicative Lagrangian. This suggests that the multiplicative Lagrangian/Hamiltonian framework offers a profound and beautiful foundation, one that reveals the underlying unity between classical and relativistic descriptions in a way that transcends traditional formulations. In essence, the multiplicative approach introduces a richer and more intricate structure to our understanding of physics, bridging the gap between different theoretical realms through a statistical perspective. 
\end{abstract}

\section{Introduction}
\begin{figure}[h]
\centering
\includegraphics[width=0.75\linewidth]{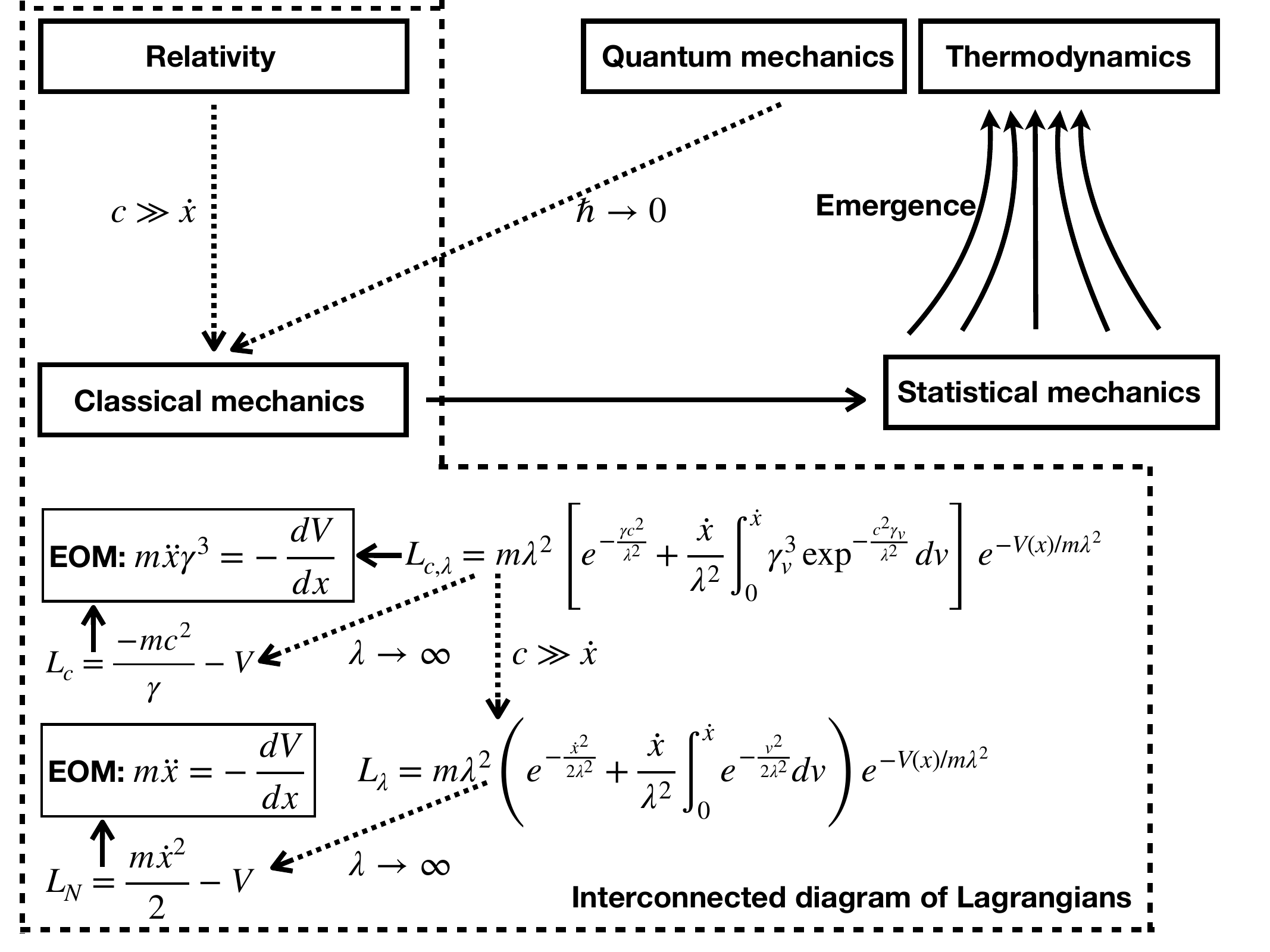}
\caption{\label{fig1} The basic hierarchical structure of physical theory.}
\end{figure}

In the landscape of physics, theoretical frameworks often exhibit a hierarchical structure: higher-level theories encompass broader phenomena and, under appropriate limits, reduce to lower-level theories. For example, Newtonian mechanics can be retrieved from special relativity in the limit of low velocities $(v\ll c )$, while classical mechanics can be recovered from quantum mechanics since Planck's constant tends to zero $(\hbar\rightarrow 0)$, see figure \ref{fig1}. This top-down viewpoint suggests that higher theories are not just corrections or extensions but provide a more complete picture, from which familiar, lower-level theories can naturally be considered as special cases. This nested structure of physical laws allows us to understand different domains of physics within a unified framework. However, not all relationships between theories are purely top-down. A striking example comes from the connection between thermodynamics and statistical mechanics. In this case, thermodynamics does not simply reduce from statistical mechanics in a formal limit. Instead, thermodynamic laws emerge from the collective behavior of a large number of microscopic degrees of freedom. Here, macroscopic concepts such as temperature and entropy are not evident at the level of individual particles, but become well-defined only through statistical averaging. This emerging perspective offers a bottom-up understanding, where a simpler phenomenological theory arises from complex underlying dynamics. This interplay between top-down derivation and bottom-up emergence provides an important conceptual backdrop for the present work. 
\\
\\
Conventionally, the relativistic Lagrangian is considered a top-down extension of the standard Lagrangian, linked through a limiting process. However, in this study, we shall explore an alternative route. We demonstrate that the relativistic Lagrangian and Hamiltonian can statistically emerge from a generalised non-relativistic formulation called the multiplicative Lagrangian \cite{MultiL}. The content of this work is as follows. Section \ref{sec2} provides a brief review of the construction of the multiplicative Lagrangian and its corresponding Hamiltonian. Section \ref{sec3} introduces a modified version of the multiplicative Lagrangian, from which the emergence of the relativistic Lagrangian and Hamiltonian is derived. Section \ref{sec4} extends the framework by including a potential term, showing that the same statistical averaging process still yields correct equations of motion, while revealing a new hierarchy of Lagrangians and Hamiltonians. Section \ref{sec5} takes a broader look at the action functional and shows that even the geometry of spacetime, whether flat or curved, can be seen as a statistical result of this approach. This suggests a bottom-up picture of how relativistic spacetime might emerge from deeper classical dynamics. Section \ref{sec6} presents the summary and conclusions, along with a discussion of open problems and potential directions for future investigation.
\section{Multiplicative Lagrangian and Hamiltonian}\label{sec2}
We begin by considering a non-standard formulation of classical mechanics in which the Lagrangian takes a \emph{multiplicative} rather than additive form \cite{MultiL}. Specifically, we assume
\begin{equation}
L(x, \dot{x}) = F(\dot{x})\, G(x),
\end{equation}
where $F(\dot{x})$ and $G(x)$ are functions to be determined. Applying the principle of stationary action,
\begin{equation}
\delta S = \delta \int_{t'}^{t''} L(x, \dot{x}) \, dt = 0,
\end{equation}
with fixed endpoints, leads to a modified Euler–Lagrange equation. Using the known equation of motion $m\ddot{x} = -dV/dx$, we find that consistency requires
\begin{align}
G(x) &= \alpha_1 e^{-A V(x)/m }, \\
F(\dot{x}) &= \alpha_2 \dot{x} - \alpha_3 \left( e^{-A \dot{x}^2 /2} + A\dot{x} \int_0^{\dot{x}} e^{-A \nu^2 /2} d\nu \right),
\end{align}
where $A$ is a constant with dimensions of inverse velocity squared, and $\alpha_1, \alpha_2, \alpha_3$ are constants. Combining these expressions, \emph{multiplicative Lagrangian} becomes
\begin{equation}\label{Lk}
L(x, \dot{x}) = \left[ k_1 \dot{x} - k_2 \left( e^{-A \dot{x}^2 /2} + A\dot{x} \int_0^{\dot{x}} e^{-A \nu^2 /2} d\nu \right) \right] e^{-A V(x)/m},
\end{equation}
where $k_1 = \alpha_1 \alpha_2$ and $k_2 = \alpha_1 \alpha_3$.
\\
\\
For the special case $k_1 = 0$, $k_2 = -m\lambda^2$, and $A = 1/\lambda^2$, this Lagrangian becomes
\begin{equation}\label{ML}
L_\lambda(x, \dot{x}) = m\lambda^2 \left( e^{-\frac{\dot{x}^2}{2\lambda^2}} + \frac{\dot{x}}{\lambda^2} \int_0^{\dot{x}} e^{-\frac{v^2}{2\lambda^2}} dv \right) e^{-V(x)/m\lambda^2}.
\end{equation}
Inserting this into the Euler–Lagrange equation yields the standard equation of motion
\begin{equation}
\left(-m\ddot{x} - \frac{dV}{dx} \right) e^{-\dot{x}^2 / 2\lambda^2} = 0,
\end{equation}
which is equivalent to Newton's law since the exponential factor is always nonzero. Furthermore, in the limit $\lambda \to \infty$, we recover the standard Newtonian Lagrangian:
\begin{equation}
\lim_{\lambda \to \infty} \left( L_\lambda - m\lambda^2 \right) = \frac{1}{2} m \dot{x}^2 - V(x) = L_N.
\end{equation}
To derive the corresponding Hamiltonian, we apply the Legendre transform:
\begin{equation}
H_\lambda(p, x) = \dot{x} \frac{\partial L_\lambda}{\partial \dot{x}} - L_\lambda.
\end{equation}
This leads to a \emph{multiplicative Hamiltonian} of the form
\begin{equation}
H_\lambda(p, x) = -m\lambda^2 e^{ -\frac{H_N}{m\lambda^2}} ,
\end{equation}
where $H_N = \frac{p^2_N}{2m} + V(x)$ is the Newtonian Hamiltonian. Once again, in the limit $\lambda \to \infty$, we recover
\begin{equation}
\lim_{\lambda \to \infty} \left( H_\lambda + m\lambda^2 \right) = H_N.
\end{equation}
Thus, both the multiplicative Lagrangian and Hamiltonian represent alternative formulations of classical mechanics that preserve the same dynamics but allow for a richer structural framework, which, as we will show, permits a natural route to relativistic mechanics.
\section{Emergence of the relativistic Lagrangian and Hamiltonian from the multiplicative formulation}\label{sec3}
In this section, we demonstrate how the relativistic Lagrangian and Hamiltonian for a free particle emerge naturally from the multiplicative formulation through an elegant integration over a parameterized family of Lagrangians\footnote{The alternative method is provided in the appendix A. However, these two approaches are equivalent.}.
\subsection{Lagrangian}
\begin{figure}[h]
\centering
\includegraphics[width=1\linewidth]{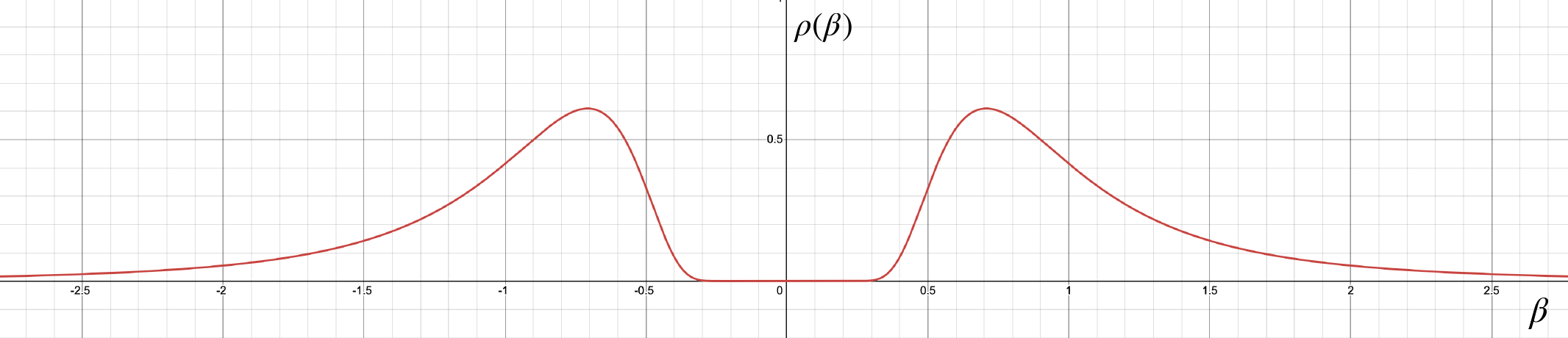}
\caption{\label{dis} The distribution function $\rho(\beta).$}
\end{figure}

In this section, we shall start with the multiplicative Lagrangian for a free particle 
\begin{equation}\label{Lf}
L_\lambda(x, \dot{x}) = -m\lambda^2 \left( e^{\frac{\dot{x}^2}{2\lambda^2}} - \frac{\dot{x}}{\lambda^2} \int_0^{\dot{x}} e^{\frac{v^2}{2\lambda^2}} dv \right) \;.
\end{equation}
We note that the parameter $\lambda^2$ is replaced by $-\lambda^2$ since we need a bound energy function $H_\lambda(p, x) = m\lambda^2 e^{ \frac{H_N}{m\lambda^2}}$. Of course, this mathematical trick does not affect the EOM. Then, we rescale the parameter such that $\lambda\rightarrow \beta\lambda$. The Lagrangian \eqref{Lf} becomes
\begin{equation}\label{Lf2}
L_{\beta\lambda}(x, \dot{x}) = -m\beta^2\lambda^2 \left( e^{\frac{\dot{x}^2}{2\beta^2\lambda^2}} - \frac{\dot{x}}{\beta^2\lambda^2} \int_0^{\dot{x}} e^{\frac{v^2}{2\beta^2\lambda^2}} dv \right) \;.
\end{equation}
Next, we introduce the distribution function, see figure \ref{dis},
\begin{equation}
    \rho(\beta)=\frac{2}{\sqrt{\pi}}\frac{e^{-\frac{1}{\beta^2}}}{\beta^4}=\frac{2}{\sqrt{\pi}}\left(\frac{H_\lambda(\dot x=0)}{H_{\beta\lambda}(\dot x=\lambda)}\right)^2
\end{equation}
which satisfies $\int_{-\infty}^{+\infty}d\beta \rho(\beta)=1$. 
We find that a statistical average of \eqref{Lf2} 
\begin{equation}
    \langle L_{\beta\lambda}\rangle_{\rho(\beta)}=\int_{-\infty}^{+\infty}d\beta \rho(\beta)L_{\beta\lambda}
\end{equation}
results
\begin{equation}
    \langle L_{\beta\lambda}\rangle_{\rho(\beta)}=-2m^2\lambda^2\sqrt{1-\frac{\dot x^2}{2\lambda^2}}\;.
\end{equation}
If we pick $\lambda^2=c^2/2$, we obtain
\begin{equation}
    L_c=\langle L_{\beta \lambda/\sqrt 2}\rangle_{\rho(\beta)}=-m^2c^2/\gamma\;,\;\; \gamma=1/\sqrt{1-\frac{\dot x^2}{c^2}}\;,
\end{equation}
which is nothing but the relativistic Lagrangian.
\subsection{Emergence of the relativistic equation of motion}
We again consider EOM with replacing $\lambda\rightarrow \beta\lambda$
\begin{equation}
e^{\dot{x}^2 / \beta^2\lambda^2} m\ddot{x} = 0\;.
\end{equation}
Then we calculate 
\begin{equation}
  \frac{1}{2}\langle e^{\dot{x}^2 / \beta^2\lambda^2} \rangle_{\rho(\beta)}m\ddot{x}= \frac{1}{2} \int_{-\infty}^{+\infty}d\beta \rho(\beta) \left(e^{\dot{x}^2 / \beta^2\lambda^2} \right)m\ddot{x}=m\ddot x{\left(1-\frac{\dot x^2}{2\lambda^2}\right)^{3/2}}=0\;.
\end{equation}
If we pick $\lambda^2=c^2/2$, we obtain the relativistic EOM: $m\ddot x\gamma^3=0$. If we introduce $d\tau=dt/\gamma$, the relativistic EOM can be compactly written as $d^2x/d\tau^2=0$. This equation means that the particle moves with constant velocity in its own proper time $\tau$, which is consistent with the fact that, in the absence of external forces, the particle moves along a geodesic in spacetime (straight-line motion).
\subsection{Hamiltonian}
We now consider the multiplicative Hamiltonian for free particle
$
    H_\lambda=m\lambda^2e^{\frac{p_N^2}{2m\lambda^2}}\;.
$
We do again replace $\lambda\rightarrow\beta\lambda$ resulting in
\begin{equation}
    H_{\beta\lambda}=m\beta^2\lambda^2e^{\frac{p_N^2}{2m\beta^2\lambda^2}}\;.
\end{equation}
If we pick $\lambda^2=c^2/2$ and consider the statistical average
\begin{equation}
    \langle H_{\beta \lambda/\sqrt 2} \rangle_{\rho(\beta)} =\int_{-\infty}^{+\infty}d\beta \rho(\beta) H_{\beta \lambda/\sqrt 2}=mc^2\gamma\;,
\end{equation}\
which is of course the relativistic Hamiltonian.
\subsection{Interpretation}
\begin{figure}[h]
\centering
\includegraphics[width=1\linewidth]{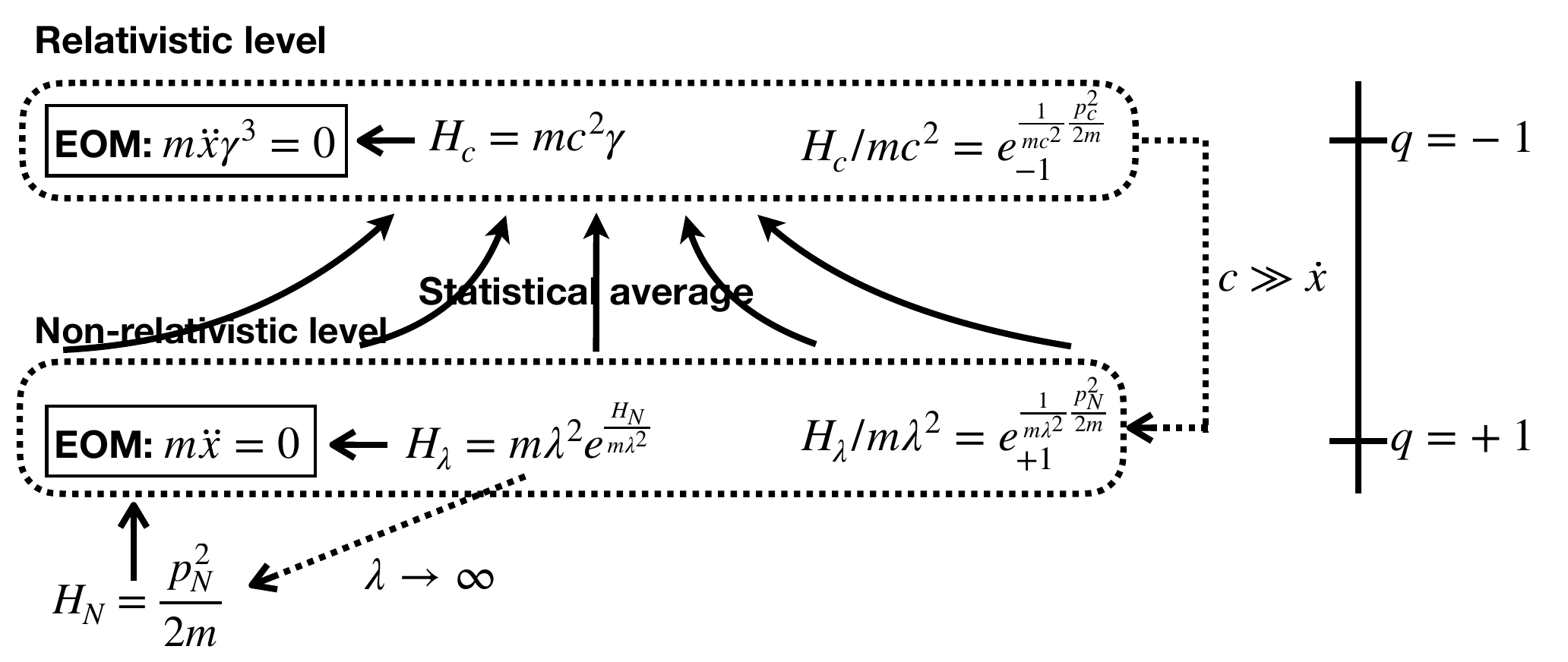}
\caption{\label{Ham} The interconnected diagram for Hamiltonians.}
\end{figure}

Recalling the relativistic Lagrangian
\begin{eqnarray}
    L_c&=&\int_{-\infty}^{+\infty}d\beta \rho(\beta)L_{\beta \lambda/\sqrt 2}=\langle L_{\beta \lambda/\sqrt 2 \rangle}\rangle_{\rho(\beta)}\nonumber\\
    &=&-mc^2/2\left[\langle \beta^2 \rangle_{\rho(\beta)}-\frac{T}{mc^2/2}\langle  1\rangle_{\rho(\beta)}-\frac{1}{2!3}\frac{T^2}{(mc^2/2)^2}\left\langle   \frac{1}{\beta^2}\right\rangle_{\rho(\beta)}+...\right]\;,
\end{eqnarray}
where $T=m\dot x^2/2$. We find that the relativistic Lagrangian is a statistical average under distribution $\rho(\beta)$ of all possible structures of the non-relativistic Lagrangian generated from the multiplicative Lagrangian.
\\
\\
Next, we consider the EOM which is
\begin{eqnarray}
    &&m\ddot x\left\langle e^{\frac{\dot x^2}{\beta^2 c^2}}\right\rangle_{\rho(\beta)}=0 \nonumber\\
    &&m\ddot x\left(\underbrace{1}_{standard \;one}+ \underbrace{\left\langle \frac{\dot x^2}{\beta^2 c^2}\right\rangle_{\rho(\beta)}+\frac{1}{2!}\left\langle \left(\frac{\dot x^2}{\beta^2 c^2}\right)^2\right\rangle_{\rho(\beta)}+...}_{non-standard\;one}\right)=0\;.
\end{eqnarray}
This equation indicates that a relativistic particle is a non-relativistic particle which explores (sum over possible) all possible relativistic fractions (weight function) $\langle (\dot x^2/\beta^2 c^2)^j\rangle_{\rho(\beta)}$, where $j=0,1,2,...$.
\\
\\
The relativistic free Hamiltonian can be expressed in terms of the $q$-deformed exponential function \cite{Tsallis2023ch3} as
\begin{equation}\label{EE1}
    H_c/mc^2 \,= e_{q}^{\frac{1}{mc^2}\frac{p_c^2}{2m}}\;,
\end{equation}
where $q=-1$. We rewrite the non-relativistic multiplicative Hamiltonian for the free particle as
\begin{equation}\label{HkA2}
    H_{\lambda}/m\lambda^2=e_q^{\frac{1}{m\lambda^2}\frac{p_N^2}{2m}}\;,
\end{equation}
where $q=+1$.  This structure was explored in \cite{krisut2024} to highlight the non-additive nature of the relativistic free Hamiltonian. However, the connection between the non-relativistic Hamiltonian $(q=+1)$ and the relativistic Hamiltonian $(q=-1)$ is not straightforward, as the procedure for upgrading the former to the latter remains nontrivial. Fortunately, through the process described in section \ref{sec2}, we gain a clear perspective on how these two Hamiltonians are interconnected from top-down and bottom-up approaches, see Figure \ref{Ham}. In other words, this work demonstrates how high-energy physics naturally emerges from low-energy physics through the proposed framework.
\\
\\
Up to this stage, we shall point out that this construction reveals that the relativistic Lagrangian (Hamiltonian) and EOM can be understood as \emph{emergent structures} which is not by imposing relativity externally, but by integrating over a parameter space of generalized classical models. This offers a novel bottom-up perspective: just as macroscopic thermodynamic laws emerge from microscopic statistical behavior, relativistic dynamics can emerge from hidden structure embedded in non-relativistic mechanics.
%
%
%
%
%
%
%
%
%
%
\subsection{Multiplicative relativistic Hamiltonian}
In \cite{MultiL}, the one-parameter version of the standard relativistic Hamiltonian was established given by
\begin{equation}
    H_{c,\lambda}=-m\lambda^2e^{-\frac{\gamma c^2}{\lambda^2}}\;
\end{equation}
Then, replacing $\lambda\rightarrow \beta\lambda$, we obtain
\begin{equation}
    H_{c,\beta\lambda}=-m\beta^2\lambda^2e^{-\frac{\gamma c^2}{\beta^2\lambda^2}}\;
\end{equation}
Next, we introduce the distribution function
\begin{equation}
    \rho(\beta)=N\frac{e^{-\frac{2c^2}{\lambda^2}\left(1-\frac{1}{\beta^2\delta}\right)}}{\beta^4}\;,\;\; 1/N=\frac{1}{4}e^{-\frac{2c^2}{\lambda^2}}\sqrt{\frac{\pi}{2}}\left(-\frac{\lambda^2\delta}{c^2}\right)^{3/2}\;,\;\;\delta=\sqrt{1-\frac{\lambda^2}{c^2}}\;.
\end{equation}
Of course, it is not difficult to see that $\int_{-\infty}^{+\infty}\rho(\beta)=1$. We proceed the same mathematical steps resulting in
\begin{equation}
    H'=\left\langle H_{c,\beta\lambda} \right\rangle =\int_{-\infty}^{+\infty}d\beta \rho(\beta) H_{c,\beta\lambda} =m\left(\frac{4c^2}{\delta}\right)\sqrt{\frac{1}{1-\dot x^2/\left(\frac{4c^2}{\delta}\right)}}\;.
\end{equation}
If we replace $4c^2/\delta\rightarrow c^2$, we obtain $ H'= mc^2\gamma$, which is a relativistic Hamiltonian. This whole calculation suggests that, with the same mathematical trick, we could not go beyond the relativistic realm.
\section{Including potential}\label{sec4}
\begin{figure}[h]
\centering
\includegraphics[width=.8\linewidth]{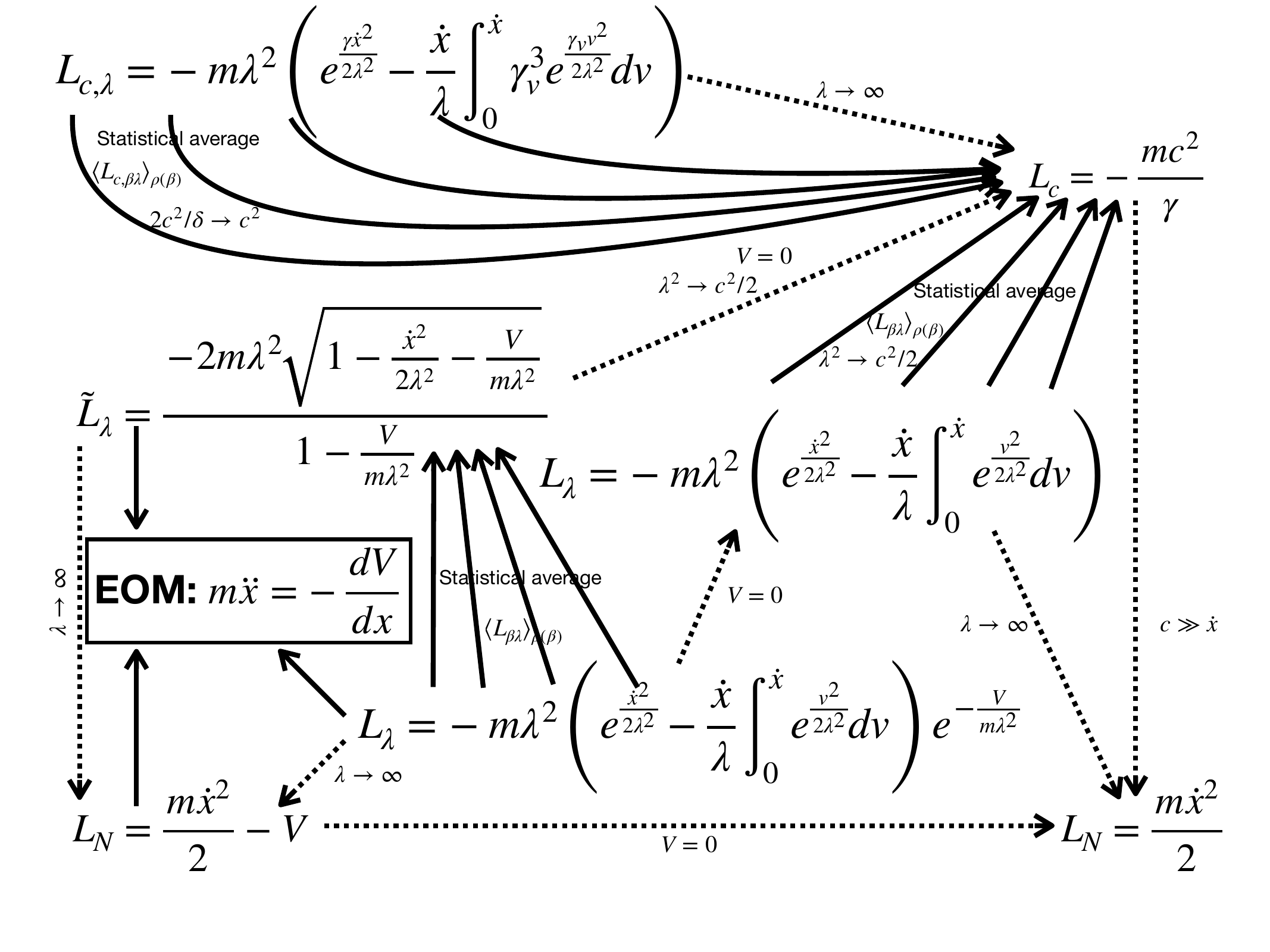}
\caption{\label{fig3} The interconnected diagram for all Lagrangians.}
\end{figure}

In this section, the full form of the multiplicative Lagrangian \eqref{ML} will be considered. We first construct the distribution function from the Hamiltonian function as follows:
\begin{equation}
    \left(\frac{H_{c,\lambda}(\dot x=0,V)}{H_{c,\beta\lambda}(\dot x=\lambda,V=0)}\right)^2=\frac{e^{-\frac{1}{\beta^2}-\frac{2V}{m\lambda^2}}}{\beta^4}\;.
\end{equation}
After normalising, we obtain
\begin{equation}
    \rho(\beta)=\frac{2}{\sqrt \pi}\frac{e^{-\frac{1}{\beta^2}}}{\beta^4}\;,
\end{equation}
which is in the same form as the previous section. Next, we compute the statistical average 
\begin{equation}\label{LN}
    \left\langle L_{\beta\lambda}\right\rangle_{\rho(\beta)}=\frac{-2m\lambda^2\sqrt{1-\frac{\dot x^2}{2\lambda^2}-\frac{V}{m\lambda^2}}}{1-\frac{V}{m\lambda^2}}\equiv \tilde L_\lambda\;,\;\;\text{where}\;, V\neq m\lambda^2\;.
\end{equation}
Applying the Legendre transformation, we obtain
\begin{equation}\label{HN}
    \tilde H_{\lambda}=\frac{2m\lambda^2}{\sqrt{1-\frac{\dot x^2}{2\lambda^2}-\frac{V}{m\lambda^2}}}\;.
\end{equation}
Using the Euler-Lagrange equation, one obtains the EOM
\begin{equation}
    \left(\frac{-2m\lambda^2}{2m\lambda^2-2V-m\dot x^2}\right)^{3/2}\left(\frac{dV}{dx}+m\ddot x\right)=0\;.
\end{equation}
Since the first bracket is not zero, we obtain the EOM: $m\ddot x=-V'$. This means that, by employing the same mathematical trick with the full multiplicative Lagrangian, we obtain a new form of Lagrangian producing the same EOM. Therefore, there exists an alternative form of the Lagrangian apart from the multiplicative Lagrangian, see figure \ref{fig3}.
\\
\\
Let us now consider the expansion of the Lagrangian \eqref{LN} with respect to the potential $V$
\begin{eqnarray}\label{LNN}
\tilde L_{\lambda}=-2m\lambda^2\sqrt{1-\frac{\dot x^2}{2\lambda^2}}\left(1+\frac{V}{m\lambda^2}\left[1-\frac{1}{2B(\dot x)}\right]+\left(\frac{V}{m\lambda^2}\right)^2\left[1-\frac{1}{2B(\dot x)}-\frac{1}{8B^2(\dot x)}\right]+... \right)\;,
\end{eqnarray}
where $B(\dot x)=1-\dot x^2/2\lambda^2$. The structure of this particular Lagrangian is very interesting. Intriguingly, if we set $V=0$ and $\lambda^2=c^2/2$, the Lagrangian \eqref{LNN} becomes $\tilde L_{\lambda}\rightarrow L_c=-mc^2/\gamma$, which is nothing but the relativistic Lagrangian. Moreover, if we consider the limit $\lambda\rightarrow \infty$, the Lagrangian \eqref{LNN} will be the non-relativistic Lagrangian $\lim_{\lambda\rightarrow\infty}(\tilde L_\lambda+2m\lambda^2)=m\dot x^2/2-V=L_N$.
\\
\\
Now, if we expand the Lagrangian with respect to $m\lambda^2$, we obtain
\begin{equation}
    \begin{aligned}\label{LH1}
\tilde{L}_\lambda+2m\lambda^2 =\   L_1 
- \frac{1}{m\lambda^2} L_2 
+ \left(\frac{1}{m\lambda^2}\right)^2 L_3 + \cdots
\end{aligned}
\end{equation}
where the first three Lagrangians in the hierarchy are given by
\begin{eqnarray}\label{lll}
    &&L_1=L_N=T-V\;,\nonumber\\
    &&L_2=\frac{1}{8} \left( T + V \right)^2 + \frac{1}{2}V \left(T + V \right) - V^2\nonumber\;,\\
    &&L_3=\frac{1}{16} \left(T + V \right)^3 + \frac{1}{8} \left( T + V \right)^2 V + \frac{1}{2} \left( T + V \right) V^2 - V^3\;.\nonumber
\end{eqnarray}
It is not difficult to see that these Lagrangians in the hierarchy \eqref{LH1} give exactly the EOM: $m\ddot x=-V'$. What we see is that this new Lagrangian \eqref{LN} gives us a way in generating a set of non-trivial Lagrangian\footnote{This new set is not the same with those obtained from the multiplicative Lagrangian \cite{MultiL}.} producing the same EOM. Therefore, this reflects again the non-uniqueness property of Lagrangian.
\\
\\
For the Hamiltonian \eqref{HN}, we find that it can be rewritten in the form
\begin{equation}
    \tilde H_\lambda=\frac{ 2m\lambda^2}{\sqrt{1-\frac{1}{m\lambda^2}(T+V)}}\;.
\end{equation}
Then, expansion with respect to $m\lambda^2$, we obtain
\begin{equation}
    \tilde H_{\lambda}+2m\lambda^2=H_1+\frac{3}{4}\frac{1}{m\lambda^2}H_2+\frac{5}{8}\left(\frac{1}{m\lambda^2}\right)^2H_3+\cdots\;,
\end{equation}
where the first three Hamiltonian in hierarchy are given by
\begin{eqnarray}
    &&H_1=T+V=H_N\;,\nonumber\\
    &&H_2= \left( T + V \right)^2=H_N^2 \nonumber\;,\\
    &&H_3= \left(T + V \right)^3=H_N^3 \;.\nonumber
\end{eqnarray}
Again, these Hamiltonians give the same EOM: $m\ddot x=-V'$. Unlike the Lagrangian structure given in \eqref{lll}, the hierarchy of the Hamiltonian $\tilde H_\lambda$ is the same as that obtained from the multiplicative Hamiltonian \cite{MultiL}, since $H_\lambda(H_N)$ and $\tilde H_\lambda(H_N)$ are functions of the standard Hamiltonian.
\section{Action and spacetime}\label{sec5}
\subsection{The 1+1 flat spacetime}
\begin{figure}[h]
\centering
\includegraphics[width=.9\linewidth]{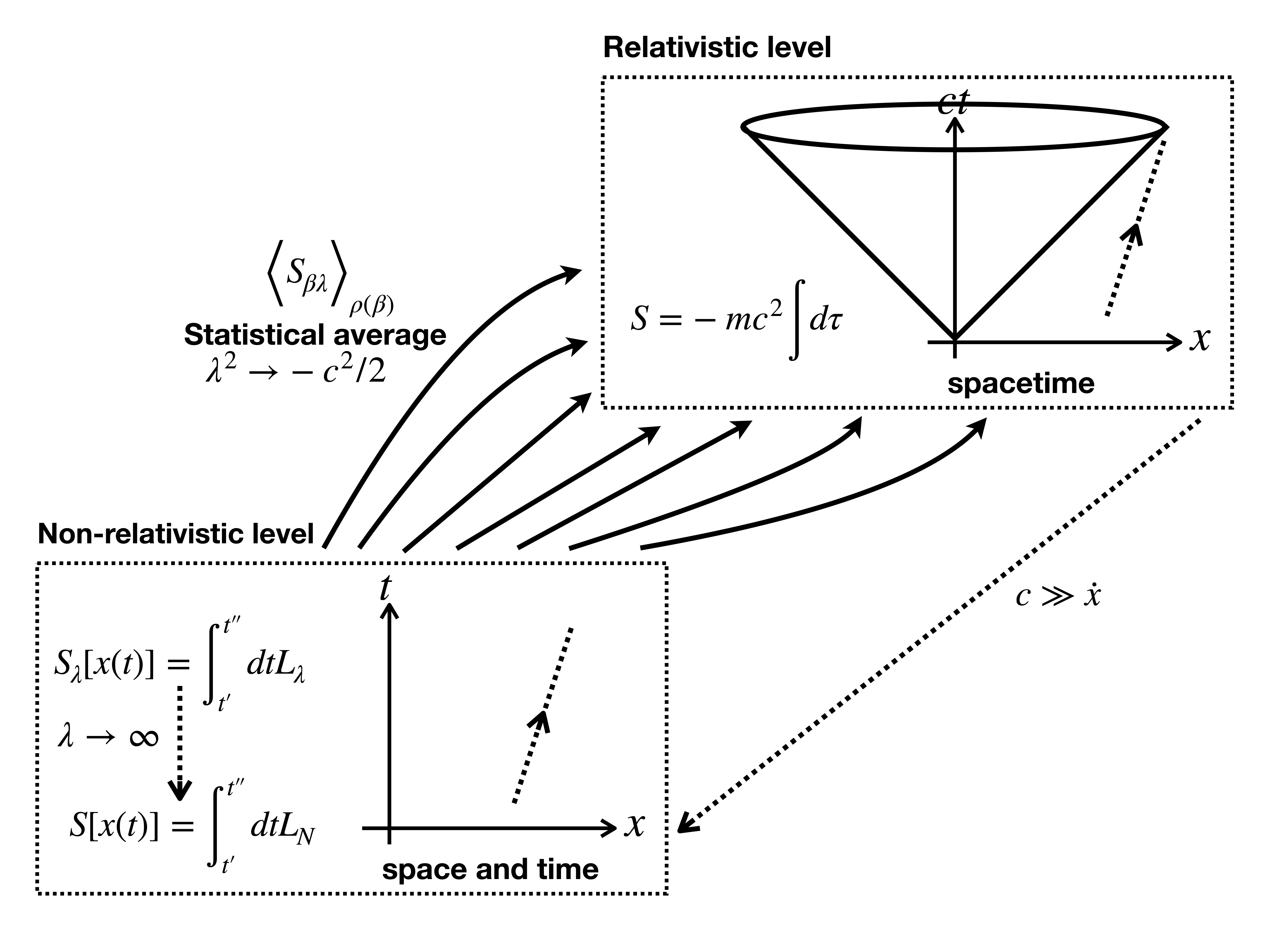}
\caption{\label{ST} Emergence of Spacetime from statistical average through the action of multiplicative Lagrangian.}
\end{figure}

In this section, we consider the action functional. We first call the action for non-relativistic free particle associated with standard Lagrangian
\begin{equation}
    S[x(t)]=\frac{m}{2}\int_{t'}^{t''}dt\dot x^2=\frac{m}{2}\int_{t'}^{t''}dtg_{ij}\dot x^i \dot x^j\;,
\end{equation}
where $g_{ij}=\delta_{ij}$ is the spatial metric in Euclidean space. Of course, the structure of spacetime is implicit: time and space are separate. Next, we shall call the action for a non-relativistic free particle associated with multiplicative Lagrangian
\begin{eqnarray}\label{AA}
    S_\lambda[x(t)]&=&-m\lambda^2\int_{t'}^{t''}dt  \left( e^{\frac{\dot{x}^2}{2\lambda^2}} - \frac{\dot{x}}{\lambda^2} \int_0^{\dot{x}} e^{\frac{v^2}{2\lambda^2}} dv \right)\nonumber\\
    &=&-m\lambda^2 \int_{t'}^{t''}dt \left(1-\frac{1}{2\lambda^2}\dot x^2 -\frac{1}{2!3}\left(\frac{1}{2\lambda^2}\right)^2\dot x^4-\frac{1}{3!5}\left(\frac{1}{2\lambda^2}\right)^3\dot x^6+...\right)\;.
\end{eqnarray}
Of course, the action \eqref{AA} gives all the terms beyond the standard formalism. 
Applying the same mathematical trick, $\lambda\rightarrow \beta\lambda$, $\lambda^2=c^2/2$ and considering the statistical average
\begin{equation}
    \left\langle S_{\beta\lambda}[x(t)]\right\rangle_{\rho(\beta)} =-mc^2\int_{t'}^{t''}dt\sqrt{1-\frac{\dot x^2}{c^2}}=-mc^2\int_{\tau '}^{\tau''}d\tau\;,
\end{equation}
where $cd\tau=ds$ and $ds^2=-d(ct)^2+dx^2$ is the interval for the $1+1$ Minkowski spacetime 
\\
\\
What we can see here is that, with the non-relativistic action \eqref{AA}, space and time are separated. Interestingly, space and time are possibly fused through the statistical average of the action with respect to the distribution $\rho(\beta)$, see figure \ref{ST}. This, once again, reflects the bottom-up perspective, where spacetime astonishingly pops up as an emergent phenomenon.
%
\subsection{The $1+1$ curved spacetime}
In this section, we shall consider the action associated with the Lagrangian \eqref{LN} given by
\begin{equation}
    S=\int_{t'}^{t''}dt\left(-mc^2 \sigma\sqrt{1-\frac{\sigma^2\dot x^2}{c^2}}\right)\;,
\end{equation}
where $\lambda^2=c^2/2$ and
\begin{equation}
    \sigma=\frac{1}{\sqrt{1-\frac{2V}{mc^2}}}\;.
\end{equation}
We now introduce the spacetime interval $s=S/mc$ resulting in 
\begin{eqnarray}
    ds&=&-dt\sigma c\sqrt{1-\frac{\sigma^2\dot x^2}{c^2}}\nonumber\;,\\
    ds^2&=&\sigma^2c^2dt^2-\sigma^4 dx^2\nonumber\;,\\
    &=&\frac{d(ct)^2}{1-\frac{2V}{mc^2}}-\frac{dx^2}{(1-\frac{2V}{mc^2})^2}\;.
\end{eqnarray}
Next, we rescale the spacetime interval as follows: $ds'^2=\sigma^{3/2}ds^2$. What we now have
\begin{equation}\label{ds0}
    ds'^2=\left(1-\frac{2V}{mc^2}\right)^{1/2}d(ct)^2-\left(1-\frac{2V}{mc^2}\right)^{-1/2}dx^2\;.
\end{equation}
This form of the spacetime interval is not standard in general relativity. Next, we shall compute the Gaussian curvature, which is an intrinsic property—it depends only on the metric, not how the surface is embedded in space\footnote{We note that in this $1+1$ spacetime case, Einstein tensor vanishes identically for any metric.}, of the spacetime \eqref{ds0}. We then compute the first and second kind Christoffel symbols
\begin{equation}
\Gamma_{\lambda\mu\nu} = \frac{1}{2} \left( \partial_\mu g_{\nu\lambda} + \partial_\nu g_{\mu\lambda} - \partial_\lambda g_{\mu\nu} \right)\;,\;\text{and}\;
\Gamma^{\lambda}_{\mu\nu} = \frac{1}{2} g^{\lambda\rho} \left( \partial_\mu g_{\nu\rho} + \partial_\nu g_{\mu\rho} - \partial_\rho g_{\mu\nu} \right)\;
\end{equation}
resulting in
\begin{equation}
    \Gamma^{1}_{00}=-\frac{V'}{2mc^2}\;,\;\Gamma^0_{10}=\Gamma^0_{01}=-\frac{V'}{2mc^2-4V}\;,\;\Gamma^1_{11}=\frac{V'}{2mc^2-4V}\;,
\end{equation}
while other components are zero. The Riemann tensor 
\begin{equation}
R^{\rho}_{\ \sigma\mu\nu} = \partial_\mu \Gamma^{\rho}_{\nu\sigma} - \partial_\nu \Gamma^{\rho}_{\mu\sigma} + \Gamma^{\rho}_{\mu\lambda} \Gamma^{\lambda}_{\nu\sigma} - \Gamma^{\rho}_{\nu\lambda} \Gamma^{\lambda}_{\mu\sigma}
\end{equation}
has non-zero component as follows
\begin{eqnarray}
    R_{1010}=-R_{0110}&=&\frac{V'^2+(mc^2-2V)V''}{2(mc^2-2V)^2}\;,\nonumber\\
    R_{1001}=-R_{0101}&=&\frac{V'^2+(mc^2-2V)V''}{2mc^2(m-2V)}\;.\nonumber\\
\end{eqnarray}
The Ricci tensor can be computed by $R^{k}_{ikj}=R_{ij}$ and we find 
\begin{eqnarray}
    R_{00}&=&-\frac{V'^2}{2mc^2(m-2V)}-\frac{V''}{2mc^2}\;,\nonumber\\
    R_{11}&=&-\frac{V'^2+(mc^2-2V)V''}{2(mc^2-2V)^2}\;.\nonumber\\
\end{eqnarray}
Therefore, the Ricci scalar is given by
\begin{equation}
    R=-\frac{\left(1-\frac{2V}{mc^2}\right)^{1/2}\left(V'^2+(mc^2-2V)V''\right)}{\left(mc^2-2V\right)^2}\;,
\end{equation}
and the curvature of the $1+1$ spacetime is $\kappa=-R/2$.
\\
\\
However, if we further consider the expansion in the limit $V\ll mc^2$ up to the first order, we obtain
\begin{equation}\label{ds1}
    ds'^2=\left(1+\frac{2\Phi}{c^2}\right)d(ct)^2-\left(1-\frac{2\Phi}{c^2}\right)dx^2\;,
\end{equation}
where $\Phi=-V/m$. Equation \eqref{ds1} reminiscently resembles the form of the spacetime interval in the limit where the particle is located far from the stationary source \cite{carroll2003spacetime}. We note here that if $1-2V/mc^2\rightarrow 0$, time component in \eqref{ds0} vanishes. This means time $t$ appears to freeze for a distant observer watching something fall in. At the same situation, if $(1-2V/mc^2)^{-1}\rightarrow \infty$, the spatial distance blows up. This suggests that the coordinate singularity(infinite values of the metric components)at $V=mc^2/2$ is not a physical singularity and, indeed, it is an artifact of this spacetime coordinate. This feature again resembles the feature that we have in the $1+1$ Schwarzchild metric. 

\section{Concluding summary}\label{sec6}
The emergence of the relativistic Lagrangian and Hamiltonian through integration over a family of non-standard Lagrangian (Hamiltonian) offers a novel perspective on how fundamental physical theories may be connected. In traditional physics, higher theories are often viewed as generalizations of lower ones: relativity generalizes Newtonian mechanics, quantum field theory generalizes quantum mechanics, and so on. In such a \emph{top-down} hierarchy, higher theories reduce to lower ones under suitable limits (e.g. $c \gg v$, $\hbar \to 0$). However, the method presented here inverts that viewpoint. By integrating on a continuous family of classical Lagrangians parameterized by $\beta$, we show that relativistic dynamics, typically seen as a fundamental theory, can instead be regarded as a \emph{emergent} phenomenon. This \emph{bottom-up} construction hints at a statistical or ensemble interpretation of relativistic motion, reminiscent of how thermodynamic laws arise from averaging over microscopic degrees of freedom in statistical mechanics. Indeed, while the classical-to-relativistic transition is usually treated as a limiting procedure, here it is shown to result from an \emph{integration over classical structures}. This suggests a new kind of structural unity: relativity emerges not only in the limit of high speeds but also from \emph{hidden layers of classical formalisms} when viewed through the lens of averaging or superposition. Moreover, at the level of the action, one finds that the structure of Einstein's spacetime naturally emerges as a statistical average of Newtonian spacetime. To the best of our knowledge, this type of connection has not been previously reported in the literature. However, several conceptual and technical aspects of this framework remain open for future investigation. In particular, the precise interpretation of time in our construction, as either an external evolution parameter or as a coordinate in an emergent spacetime, requires deeper clarification. Furthermore, the role and uniqueness of the chosen distribution function $\rho(\beta)$ warrants further analysis. While our selected form yields the desired relativistic dynamics, it remains to be studied whether this choice is unique, whether alternative distributions lead to physically meaningful deformations, or whether underlying principles might determine $\rho(\beta)$ naturally. Here are some final remarks. This approach would possibly echo ideas in other fields\footnote{In gravity, space time geometry might emerge from quantum entanglement and in condensed matter, the emergence of superconducting states in cuprates and iron-based materials remains an unsolved problem in condensed matter physics.}. Similarly, our result suggests that relativistic dynamics might not be fundamental but rather \emph{collective}, \emph{coarse-grained}, or \emph{ensemble-averaged} behavior of deeper, non-standard classical systems. We observe that the multiplicative Hamiltonian and Lagrangian formulations, as introduced in \cite{MultiL}, may serve as alternative frameworks that enrich our understanding of physical theories. Although still in an exploratory stage, these formulations have the potential to uncover subtleties that might not show up in standard approaches, and thus merit further theoretical investigation, see \cite{krisut2024,BUKAEW202357,Supanyo2022,Supanyo:2023jkh, Supanyo2024,supanyo2025}.


\appendix
\section*{Appendix A} \label{Apen}
\section{Alternative method}
In this appendix, the alternative of statistical average of the non-relativistic multiplicative Lagrangian and Hamiltonian will be discussed.
\subsection{Lagrangian}

We begin with the multiplicative Lagrangian \eqref{Lk} in the form
\begin{equation}
L_{k_2, A}(x, \dot{x}) = -k_2 \left( e^{-A \dot{x}^2 /2} + A\dot{x} \int_0^{\dot{x}} e^{-A \nu^2 /2} \, d\nu \right).
\end{equation}
To make contact with relativistic dynamics, we reparametrise
\begin{equation}
k_2 = -K e^{-\alpha^2}, \qquad A = -\frac{2\alpha^2}{c^2},
\end{equation}
where $K$ is a constant with units of energy, $\alpha \in \mathbb{R}$, and $c$ is the speed of light. The Lagrangian then becomes
\begin{equation}\label{LKK}
L_{K,\alpha,c}(x, \dot{x}) = K \left( e^{-\alpha^2 (1 - \dot{x}^2 / c^2)} - \frac{2\alpha^2 \dot{x}}{c^2} \int_0^{\dot{x}} e^{-\alpha^2 (1 - \nu^2 / c^2)} \, d\nu \right).
\end{equation}
Remarkably, for any fixed $\alpha$, this Lagrangian yields the same classical equation of motion $m\ddot{x} = -V'$. The key insight is to treat the family of these Lagrangians as forming a Gaussian-like distribution over $\alpha$ and integrate over all values yielding
\begin{equation}
\int_{-\infty}^{\infty} L_{K, \alpha, c} \, d\alpha = K\sqrt{\pi} \, \gamma,
\end{equation}
where $\gamma = \left(1 - \dot{x}^2 / c^2\right)^{-1/2}$ is the Lorentz factor. By choosing $K\sqrt{\pi} = -mc^2$, we recover the free-particle relativistic Lagrangian: $L_c = -mc^2 /\gamma$.

\subsection{Emergence of the relativistic equation of motion}
We can verify this result by examining the equations of motion. The multiplicative EOM is
\begin{equation}
e^{-\alpha^2 \dot{x}^2 / c^2}m \ddot{x} = 0\;,
\end{equation}
which can be rewritten as
\begin{equation}
\frac{d}{dt} \left(Ke^{-\alpha^2} e^{-\alpha^2 \dot{x}^2 / c^2} m\dot{x} \right) = 0.
\end{equation}
Integrating over $\alpha$, we obtain
\begin{equation}
 \frac{d}{dt} (m\gamma \dot{x}) = 0\;,
\end{equation}
which is the relativistic EOM of the free particle.
\subsection{Hamiltonian}
A similar result holds in the Hamiltonian formulation. Consider the multiplicative Hamiltonian:
\begin{equation}
H_{k, A} = k e^{\ -A \frac{p_N^2}{2m} }, \qquad \text{with } p_N = m\dot{x}.
\end{equation}
Using the substitutions
\begin{equation}
k = -K e^{-\alpha^2}, \qquad A = -\frac{2\alpha^2}{m^2 c^2},
\end{equation}
we obtain
\begin{equation}\label{HKK}
H_{K,\alpha,c} = -K e^{ -\alpha^2 \left( 1 - \frac{p_N^2}{m^2 c^2} \right) }.
\end{equation}
Integrating over $\alpha$
\begin{equation}
\int_{-\infty}^{\infty} H_{K, \alpha, c} \, d\alpha = -K \sqrt{\pi} \, \gamma.
\end{equation}
Again, choosing $K\sqrt{\pi} = -mc^2$, we recover the relativistic Hamiltonian: $H_c = \gamma mc^2$.
\subsection{Statistical average}
\begin{figure}[h]
\centering
\includegraphics[width=1\linewidth]{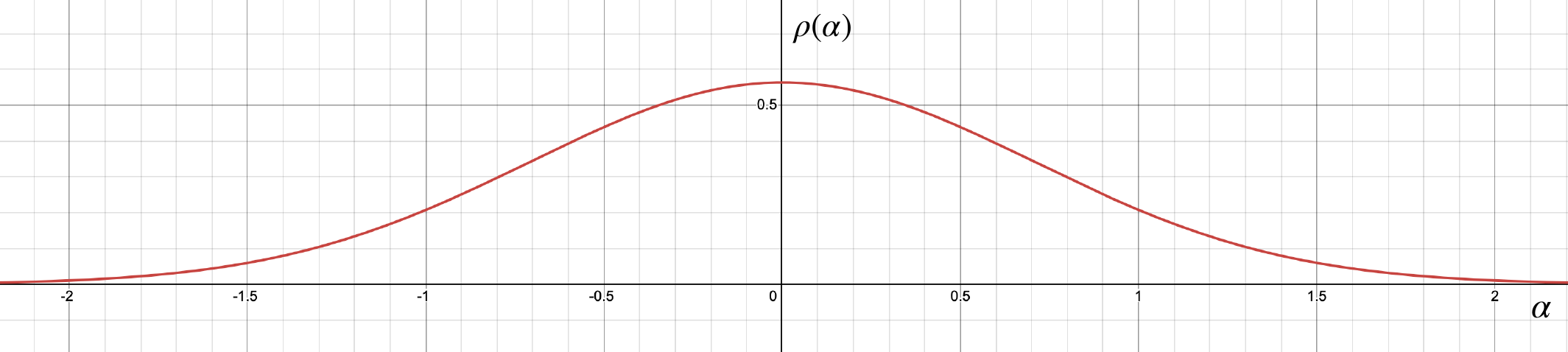}
\caption{\label{dis2} The distribution function $\rho(\alpha)$.}
\end{figure}

Again, with this different mathematical trick, we still capture the process as the statistical average of the non-standard Lagrangian. Recalling \eqref{LKK}, we have
\begin{equation}
    L_c=\int_{-\infty}^{+\infty}d\alpha\rho(\alpha)\left(-mc^2\left[e^{-\frac{\alpha^2\dot x^2}{c^2}}-\frac{2\alpha^2\dot x}{c^2}\int_{0}^{\dot x}e^{-\frac{\alpha^2v^2}{c^2}}dv\right]\right)\;,
\end{equation}
where $\rho(\alpha)=e^{-\alpha^2}/\sqrt{\pi}$, see figure \ref{dis2}. We note that the Hamiltonian \eqref{HKK} can be re-expressed into the statistical average as well.

\section*{Acknowledgement}
We gratefully acknowledge Amorthep Tita for his assistance with the curvature calculations. We would also like to express our sincere gratitude to Lunchakorn Tannukij for the valuable and insightful discussions. This research has received funding support from the NSRF through the Human Resources Program Management Unit for Human Resources \& Institutional Development, Research and Innovation [grant number 39G670016]. We acknowledge the support from the Petchra Prajomklao Ph.D. Research Scholarship from King Mongkut’s University of Technology Thonburi (KMUTT).

\bibliographystyle{unsrt}
\bibliography{references}

\end{document}